\documentclass[preprint,prfluids,nofootinbib,floats,aps,amsmath,amssymb]{revtex4-2}

\usepackage{graphicx}%
\usepackage{subcaption}
\usepackage{multirow}%
\usepackage{mathrsfs}%
\usepackage{xcolor}%
\usepackage{textcomp}%
\usepackage{booktabs}%
\usepackage{overpic}
\usepackage[normalem]{ulem}
\usepackage{bm}

\newcommand{\mean}[1]{\left\langle#1\right\rangle}

\begin{document}

\title{Lagrangian chaos and the enstrophy cascade in Ekman-Navier-Stokes two-dimensional turbulence}

\author{F.M. Ventrella{\footnote{Corresponding author: francescomichele.ventrella@unito.it}}, 
V.J. Valad\~{a}o, G. Boffetta, S. Musacchio, F. De Lillo}
\affiliation{Dipartimento di Fisica and INFN - Università degli Studi di Torino, Via Pietro Giuria, 1, 10125 Torino TO, Italy}

\begin{abstract}
Two-dimensional turbulence with linear (Ekman) friction exhibits spectral properties that deviate from the classical Kraichnan prediction for the direct enstrophy cascade. In particular, for sufficiently small viscosity and large friction, the enstrophy flux is suppressed in the cascade and, as a consequence, the small-scale vorticity field becomes passively transported by the large-scale, chaotic flow. We numerically address this problem by investigating how the statistics of the Lagrangian Finite Time Lyapunov Exponent in 2D Ekman-Navier-Stokes simulations are affected by the friction coefficient and by the other parameters of the flow. We derive a simple phenomenological model that interpolates the dependence of the Lyapunov exponent on the flow statistics from the large friction limit, where analytical predictions are available, to the small friction region. We find that the distribution of the FTLE around this mean value is always close to a Gaussian, and this allows to make a simple prediction for the correction of the spectral slope of the direct cascade which is in very good agreement with the numerical results. 
\end{abstract}

\maketitle
\newpage

\section{Introduction}

The distinctive feature of 2D turbulence lies in the existence of two quadratic inviscid invariants, energy and enstrophy, which leads to a radically different phenomenology compared to classical three-dimensional (3D) turbulence \cite{Kraichnan1967,Boffetta2012}. In 2D turbulence, energy flows toward large scales, producing an inverse cascade, while enstrophy is transported by a direct cascade toward the small scales. The statistical properties of the two cascades have been studied in detail both numerically \cite{Frisch1984,Boffetta2010} and experimentally \cite{Zhu2023,Paret1997,Rivera1998} and 2D Navier-Stokes turbulence has become a prototype for two-dimensional turbulent flows and, more in general, for other systems that exhibit a double cascade, such as wave turbulence \cite{zhu2023direct,korotkevich2008simultaneous,nazarenko2011wave}. 

The interaction of the two-dimensional flow with the three-dimensional environment introduces additional terms to the 2D Navier-Stokes equation which can be often represented by a linear dissipation. This is the case of the bottom friction induced by the container in a thin layer of fluid \cite{Boffetta2005}, the air friction on flowing soap film \cite{Rivera2000} or the Ekman friction in geophysical flows \cite{Vallis2017}. Linear friction has also the important role of removing the energy transferred at large scales by the inverse cascade and thus to allow the system to reach a statistically stationary state \cite{Boffetta2012}.

The effect of friction on the direct cascade is more subtle, since it removes the enstrophy at all scales and thus produces a cascade with a non-constant enstrophy flux. One important consequence is the correction of the spectral slope in the cascade which becomes steeper than the dimensional prediction. Moreover, in these conditions the vorticity at small-scale can be considered passive and one can exploit the results obtained for the statistics of a passive scalar with finite lifetime transported by a smooth flow \cite{Chertkov1998,Nam1999}. On this basis, one expects that the vorticity statistics can be expressed in terms of the friction coefficient and the Lagrangian statistics of stretching rates commonly characterized in terms of Finite Time Lyapunov Exponents (FTLE) \cite{Nam2000,Boffetta2002}.

In this work we investigate in detail the effects of linear friction on the distribution of the Lagrangian FTLE and the consequences on the statistics of the direct cascade. Our study is based on high-resolution numerical simulations of the Ekman–Navier–Stokes equations together with long-time integration of Lagrangian trajectories to extract the FTLE statistics. 
In particular, we study the dependence of the Lyapunov exponent on the parameters of the flow and we are able to derive a simple expression which reproduces the numerical data in a wide range of friction coefficient. Moreover, from a Gaussian model for the distribution of the FTLE, we obtain a prediction for the spectral correction of the direct cascade in close agreement with the values measured in the simulations. 

The remaining of this paper is organized as follows: Section~\ref{sec:eqs} reviews the phenomenology of 2D turbulence with Ekman friction and revisits the connection between energy spectrum and the statistics of Lagrangian trajectories while
Section~\ref{sec:model_lyap} reports theoretical predictions for the Lyapunov exponents.
Section~\ref{sec:results} presents the numerical setup and the numerical results, both for the statistics of the Lyapunov exponents and the spectrum in the direct cascade.  
Finally, Section~\ref{sec:concl} summarizes our findings and discusses the perspectives for future work.

\section{Theoretical Background}
\label{sec:eqs}

\subsection{Phenomenology of 2D turbulence with friction}
\label{sec:eqs_2d}
We consider the 2D Navier-Stokes equations written for the scalar vorticity field
$\omega(\mathbf{x},t)=\mathbf{\nabla} \times \mathbf{u}(\mathbf{x},t)$
\begin{equation}\label{eq1}
\frac{\partial \omega}{\partial t} + \mathbf{u}\cdot\nabla \omega 
= \nu \nabla^2 \omega - \alpha \omega + f_\omega ,
\end{equation}
where $\nu$ is the viscosity, $\alpha$ the friction coefficient and $f_\omega=\mathbf{\nabla} \times \mathbf{F}$ represents an external forcing. In the absence of forcing and dissipation, equation (\ref{eq1}) conserves the kinetic energy $E=\mean{|{\mathbf u}|^2}_L/2$ and the enstrophy $Z=\mean{\omega^2}_L/2$, where   $\mean{(.)}_L$ indicates the spatial average on the doubly-periodic square domain of size $L$. When $\nu\neq0$, $\alpha\neq0$, and $f_\omega\neq0$, the balance equations for energy and enstrophy in statistically stationary equilibrium read
\begin{equation}
\frac{dE}{dt} = -2 \nu Z - 2 \alpha E + \mean{{\mathbf u} \cdot {\mathbf F}}_L =
-\varepsilon_{\nu}-\varepsilon_{\alpha}+\varepsilon_I =0\ ,\
\label{eq2}
\end{equation}
and
\begin{equation}
\frac{dZ}{dt} = -2 \nu P - 2 \alpha Z + \mean{\omega f_\omega}_L =
-\eta_{\nu}-\eta_{\alpha}+\eta_I =0\ ,\
\label{eq3}
\end{equation}
where 
$\varepsilon_I = \mean{{\mathbf u} \cdot {\mathbf F}}_L$ and 
$\eta_I = \mean{\omega f_\omega}_L$ are the energy and enstrophy injection rates,
$\varepsilon_\nu = 2 \nu Z$ and $\varepsilon_\alpha = 2 \alpha E$ are the energy dissipation rates due to viscosity and friction, 
$\eta_\nu = 2 \nu P$ and $\eta_\alpha = 2 \alpha Z$ are the enstrophy dissipation rates due to viscosity and friction,
and $P=\mean{|\mathbf\nabla\times(\omega\mathbf{e}_z)|^2}_L/2$ is the palinstrophy. 

We assume that the forcing is localized around a characteristic injection scale $\ell_I=2\pi\sqrt{\varepsilon_I/\eta_I}$, while viscosity and friction dissipation define a small viscous scale $\ell_{\nu}=2\pi\sqrt{\varepsilon_{\nu}/\eta_{\nu}}$ and a large friction scale $\ell_{\alpha}=2\pi\sqrt{\varepsilon_{\alpha}/\eta_{\alpha}}$. A scale separation, $\ell_{\nu} \ll \ell_I \ll \ell_{\alpha}$, is achieved in the limit of small viscosity and friction coefficient ($\nu\to 0$, $\alpha \to 0$). In these conditions, the dynamics of 2D turbulent flows display the development of a direct enstrophy cascade, characterized by a constant flux of enstrophy along the scales $\ell_{\nu} \ll \ell \ll \ell_I$ and an inverse energy cascade with a constant energy flux in the scales $\ell_I \ll \ell \ll \ell_{\alpha}$ \cite{Kraichnan1967,Boffetta2012}. 

The assumption of a constant enstrophy flux in the range of scales of the direct cascade leads to the prediction of a power law energy spectrum $E(k)\sim k^{-3}$ with possible logarithmic corrections \cite{Kraichnan1967,Kraichnan1971}. 
However, numerical and experimental studies \cite{Nam2000,Bernard2000,Boffetta2002,Boffetta2005,Valadao2025} have shown that the presence of friction causes the steepening of the energy spectrum $E(k)\sim k^{-(3+\xi)}$, with a friction-dependent correction to the spectral slope $\xi(\alpha) >0$.  
In these conditions, vorticity fluctuations at small scales can be considered as being transported like a passive scalar by a smooth, large-scale, chaotic velocity field. The combined effect of chaotic advection and linear damping causes anomalous scaling and small-scale intermittency in the statistics of the vorticity field \cite{Nam2000,Bernard2000}.

The phenomenon of the steepening of the spectrum can be interpreted as a result of the dissipation of enstrophy by friction, which causes a reduction of the enstrophy flux in the direct cascade. In the limit of strong friction ($1/\alpha \to 0$), an estimate of the residual dissipation of enstrophy due to viscous forces at small scales can be obtained by means of a classic argument by Fj{\o}rtoft\cite{Fjortoft1953}. Using the dimensional estimate for the palinstrophy $P\sim Z/\ell^2_\nu$, the viscous dissipation rate of enstrophy can be written as follows: 
\begin{equation}
    \eta_\nu = 2\nu P \sim 2\nu \frac{Z}{\ell_\nu^2}
    = \frac{\nu}{\alpha \ell_\nu^2}\eta_\alpha \ . \
    \label{eq_lim1}
\end{equation} 
Inserting this expression in the energy balance~\eqref{eq2} by assuming the dimensional scaling relation $\varepsilon_\ell\sim\ell^2\eta_\ell$, one obtains: 
\begin{equation}
    \frac{\eta_\nu}{\eta_I}=1-\frac{\ell_I^2}{\ell_\alpha^2+\frac{\nu}{\alpha}}
    \simeq \frac{\nu}{\alpha\ell_I^2} + O\left(\frac{1}{\alpha^2}\right),
    \label{eq_lim2}
\end{equation}
where we assumed $\ell_\alpha\simeq \ell_I$ for sufficiently large $\alpha$. The latter assumption is justified by the fact that in the presence of intense friction, the energy is dissipated at the injection scale, preventing the development of the energy cascade at scales $\ell > \ell_I$. 

From the enstrophy balance~\eqref{eq3} and \eqref{eq_lim1} one has: 
\begin{equation}
    \frac{\eta_\nu}{\eta_I} 
    = \frac{\nu}{\alpha \ell_\nu^2}\left(1 - \frac{\eta_\nu}{\eta_I}\right) 
    \simeq \frac{\nu}{\alpha \ell_\nu^2} + O\left(\frac{1}{\alpha^2}\right)\;.
    \label{eq_lim3}
\end{equation}
The comparison between \eqref{eq_lim2} and \eqref{eq_lim3} shows that 
 in the limit of large $\alpha$, the viscous dissipation of enstrophy occurs at the forcing scale $(\ell_\nu \simeq \ell_I)$. 
 This is indeed consistent with the spectral definition of viscous dissipation rate,
\begin{equation}
    \eta_\nu = 2\nu \int_{k_{\min}}^{k_{\max}} E(k)k^4\,dk
    \sim \int_{k_{\min}}^{k_{\max}} k^{1-\xi}\,dk,
    \label{eq_lim4}
\end{equation}
where we used the power-law scaling of the spectrum $E(k) \sim k^{-(3+\xi)}$.  
The integral in Eq.~\eqref{eq_lim4} is dominated by the infrared cutoff $k_{\min}$ (i.e., by the large scale $\ell_I$) provided that the friction coefficient $\alpha$ is large enough to produce a spectral correction $\xi > 2$, as discussed in the following Section.

Equation~\eqref{eq_lim2} shows that the viscous enstrophy dissipation $\eta_\nu$ 
vanishes in the limit $1/\alpha \to 0$. 
In this limit, the enstrophy balance \eqref{eq3} is dominated by the friction dissipation 
$\eta_I = \eta_\alpha = 2\alpha Z$, and the enstrophy is inversely proportional to the friction coefficient 
\begin{equation}
Z =  \frac{\eta_I}{2\alpha}\;.
\label{eq_lim5}
\end{equation}

\subsection{Finite Time Lyapunov Exponents and spectral correction}
\label{sec:eqs_Cramer}

The Lagrangian chaoticity of a smooth flow is usually quantified by the FTLE, defined as 
\begin{equation}
\gamma_T = \lim_{|\delta \mathbf x(0)|\rightarrow 0}\frac{1}{T} \log \left( \frac{|\delta \mathbf x(T)|}{|\delta \mathbf x(0)|} \right),
\label{gammat}
\end{equation}
where $\delta\mathbf{x}(t)=\mathbf{x}_2-\mathbf{x}_1$ is the separation between two infinitesimally close Lagrangian particles. The Lagrangian Lyapunov exponent $\lambda$ of the flow is obtained as the limit
\begin{equation}
    \lambda = \lim_{T \rightarrow \infty} \gamma_T.
    \label{linf}
\end{equation}
For finite $T$, the FTLE $\gamma_T$ follows a probability distribution 
$P(\gamma_T)$ which, for large $T$, can be expressed in the large deviations form
\begin{equation}
P(\gamma_T) \sim e^{-T C(\gamma_T-\lambda)},
\label{LD}
\end{equation}
where, in general, the Cram\'er function $C(x)$ is convex ($C''(x) > 0$) and has a minimum in $x=0$, where $C'(0) = C(0) = 0$.
These properties guarantee that $\lim_{T\to\infty}P(\gamma_T)=\delta(\gamma_T-\lambda)$
and that $\mean{\gamma_T}=\lambda$, where $\mean{(.)}$ represents the average over the distribution $P(\gamma_T)$.

The simplest model for the Cram\'er function is a quadratic form
\begin{equation}
C(x) = \frac{a x^2}{2}, 
\label{C2}  
\end{equation}
which corresponds to a Gaussian probability distribution (\ref{LD}) and $a$ is given by the second moment of this distribution
\begin{align}
\lim_{T\rightarrow \infty} T\mean{\left(\gamma_T-\lambda\right)^2} &=\cfrac{1}{a} \ .\ 
\label{limit2}
\end{align} 

A higher order, cubic approximation to the Cram\'er function allows for taking into account an asymmetry in the distribution of the FTLE
\begin{equation}
C(x) = \frac{a x^2}{2}+\frac{bx^3}{3},
\label{C3}
\end{equation}
where the parameter $b$ is related to the third moment of the distribution
\begin{equation}
   \lim_{T\rightarrow \infty} T^2\mean{\left(\gamma_T-\lambda\right)^3}=\cfrac{2b}{a^3} \ .\ 
\label{limit3}
\end{equation} 
We remark that (\ref{C3}) has to be considered as an approximation of the Cram\'er function around its minimum, since it does not produce a well-behaved distribution at large negative arguments.

According to the analogy between the dynamics of vorticity fluctuations at small scales in the presence of friction and the chaotic advection of a passive scalar field with finite lifetime proposed in \cite{Nam2000,Bernard2000}, the correction $\xi$ to the slope of the energy spectrum in the direct cascade is related to the statistics of FTLE in terms of the Cram\'er function as
\begin{equation}
\xi=\min_\gamma \left\{ \cfrac{C(\gamma-\lambda) + 2\alpha}            {\gamma} \right\}\ .\
\label{chi}
\end{equation}
In the absence of fluctuations of the Lyapunov exponent, we have $P(\gamma_T)=\delta(\gamma_T-\lambda)$ and Eq.~(\ref{chi}) becomes $\xi_{(0)}=2 \alpha/\lambda$. A refined explicit expression for the correction is obtained by using the quadratic approximation~(\ref{C2}) of the Cram\'er function in 
(\ref{chi}) which gives
\begin{equation}
\xi_{(2)}=a\lambda\left( \sqrt{1+\cfrac{4\alpha}{a\lambda^2}} -1 \right) \ .\
\label{chi2}
\end{equation}
A further refinement of the spectral correction $\xi_{(3)}$ can be obtained from the cubic 
approximation~\eqref{C3} of the Cram\'er function which produces a complex explicit expression. 

\subsection{Predictions for the Lyapunov exponent}
\label{sec:model_lyap}


In the high-friction regime, it is possible to derive a prediction for the Lyapunov exponent and for the statistics of the FTLE in terms of the friction coefficient $\alpha$ and the enstrophy injection rate $\eta_I$. In this limit, the advection and viscous terms in the Ekman-Navier-Stokes equation~(\ref{eq1}) become negligible~\cite{valadao2025high} and the dynamics of the large-scale vorticity is dominated by friction and forcing terms. In the case of Gaussian random forcing, the dynamics can be traced back to an Ornstein-Uhlenbeck process with a correlation time given by the inverse of the friction coefficient. The resulting vorticity field is a short-correlated in time, Gaussian field with a temporal correlation function given by
\begin{equation}
\label{eq:vort_corr}
\langle \omega(t)  \omega(t') \rangle_G = 2 Z e^{-\alpha|t-t'|} \;,
\end{equation}
where $\langle (\cdot)\rangle_G$ denotes the average on the Gaussian statistics. 

In the limit $1/\alpha \to 0$ one recovers the Kraichnan ensemble~\cite{kraichnan1968small}, i.e., a white-in-time, Gaussian field with temporal correlation function given by~\cite{valadao2025high} 
\begin{equation}
\label{eq:vort_corr_kraichnan}
\langle \omega(t)  \omega(t') \rangle_G = 16 D_1\delta(t-t') \;,
\end{equation}
where $D_1$ is a constant related to two-particle dispersion~\cite{falkovich2001particles}. By matching the time integral of \eqref{eq:vort_corr} with that of \eqref{eq:vort_corr_kraichnan}, one gets the relations $D_1 = Z/4\alpha = \eta_I/8\alpha^2 = Z^2/2\eta_I$, where the last two equalities follow from \eqref{eq_lim5}. 

In the framework of the Kraichnan model, the statistics of the FTLE are completely determined by the parameter $D_1$.  
In particular, the Cram\'er function of a two-dimensional, incompressible, spatially smooth velocity field, with Gaussian statistics and white-in-time correlations, is given by the quadratic form 
\begin{equation}
\label{eq:Cramer_kraichnan}
C(\gamma-\lambda) = \frac{(\gamma-\lambda)^2}{2\lambda} \ ,\ 
\end{equation}
where $\lambda=2D_1$ \cite{falkovich2001particles}. This leads to the prediction for the Lyapunov exponent in the high-friction asymptotics of the Ekman-Navier-Stokes system: 
\begin{equation}
\label{eq:lyap_kraichnan}
\lambda = \frac{Z}{2\alpha} =\frac{\eta_I}{4\alpha^2} = \frac{Z^2}{\eta_I} \ .\
\end{equation}
The comparison between \eqref{eq:Cramer_kraichnan} and the cubic approximation~\eqref{C3} 
shows that, in the regime of strong friction $\alpha \eta_I^{-1/3} \gg 1$, one expects $a\lambda \simeq 1$ and $b \simeq 0$.
Moreover, using \eqref{eq:lyap_kraichnan} and \eqref{chi2}, one could also obtain the prediction for the spectral correction in the same limit.
It is worth noting that the predictions \eqref{eq:lyap_kraichnan} does not contain any free parameter. In other words, in the high-friction regime, the statistics of the FTLE are completely expressed in terms of the ratio between the enstrophy, injection rate, and the friction coefficient, quantities which can be easily measured.

Equation~\eqref{eq:lyap_kraichnan} has a simple physical interpretation. The value of the Lyapunov exponent is determined by the intensity of the velocity gradients and by their correlation time. Considering that $Z \propto \langle |\nabla {\bm u} |^2\rangle_L$ and that in the high-friction limit the correlation time is $\tau = 1/\alpha$, one gets $\lambda \propto \tau Z = Z/\alpha$. By using \eqref{eq_lim5}, one can rewrite this relation as $\lambda \propto Z^2/\eta_I$. This argument suggests the possibility of deriving a scaling prediction also for the low-friction regime $\alpha \eta_I^{-1/3} \ll 1$. In this case, the correlation time is determined by the advection time scale. A simple dimensional estimate can be given by $\tau = 1/\sqrt{Z}$, which leads to $\lambda \propto \tau Z \propto \sqrt{Z}$. 

The scaling of the Lyapunov exponent, both in the limit of high and low friction, can be matched by a general model, $\lambda \propto \tau Z$, in which $\tau$ is a combination of all the time scales associated with the friction ($\alpha$), the advection ($Z$), and the enstrophy injection ($\eta_I$). 
A suitable choice is $\tau = (AZ^{1/2} + B\eta_I^{1/3} + C \eta_I^{2/3}Z^{-1/2} + 2\alpha)^{-1}$, 
i.e., a linear combination of the inverse times, which is dominated by the fastest time scale in the system. 
This leads to the following model: 
\begin{equation}
\label{model_lyap}
\lambda = \eta_I^{1/3} \frac{\Omega^4}{A\Omega^3+B\Omega^2+C\Omega+1}
\end{equation}
where $\Omega = \sqrt{Z}\eta_I^{-1/3}$ is the dimensionless rms vorticity and $A$, $B$, $C$ are numerical coefficients to be determined and we used (\ref{eq_lim5}) in the large friction limit. The degree of the polynomial in the denominator of Eq.~\eqref{model_lyap} is dictated by the scaling of two asymptotic regimes $\lambda \propto \Omega $ (low-friction i.e. high-enstrophy) and $\lambda \propto \Omega^4 $ (high-friction i.e. low-enstrophy). Indeed, for $\Omega \ll 1$ one recovers the prediction of the Kraichnan model $\lambda = Z^2/\eta_I$, while for $\Omega \gg 1$ one gets $\lambda = A \sqrt{Z} - \eta_I^{1/3}B/A^2 + O\left(Z^{-1/2}\right)$. 

\section{Numerical simulations and results}
\label{sec:results}

Direct numerical simulations of Eq.~\eqref{eq1} have been performed by means of a pseudo-spectral code with a fourth-order Runge--Kutta time scheme implemented on GPUs. A detailed description of the code is reported in \cite{Valadao2025}. The computational domain is a square of size $L_x = L_y = 2\pi$ with periodic boundary conditions. It is discretized with a uniform grid of resolution $N = N_x = N_y=1024$. The flow is sustained by a Gaussian random forcing with zero mean and white-in-time correlations, acting within a narrow spherical shell of thickness $\delta k$ centered at $k_I$ in wavenumber space. This forcing protocol injects energy and enstrophy at constant rates $\varepsilon_I$ and $\eta_I$, respectively, which are related as $\eta_I \approx \varepsilon_I k_I^2$ for $k_I \gg \delta k$. We performed a set of 25 different simulations, varying the friction coefficient $\alpha$ while keeping the other parameters constant. The numerical values of the simulation parameters are reported in Table~\ref{tablegen}.

Together with the Navier-Stokes equation, for each simulation, we also integrated an ensemble of $N_L=100$ Lagrangian trajectories starting from initial random positions. The numerical integration of the trajectories $\dot{\mathbf{x}} = \mathbf{u} (\mathbf{x},t)$ is performed via a fourth-order Runge-Kutta scheme with linear interpolation of the fluid velocity field at the position $\mathbf{x}$. The computation of the FTLE $\gamma_T$ is performed according to standard methods~\cite{Benettin1980} by following along each Lagrangian trajectory the time evolution of an infinitesimal displacement $\delta\mathbf{x}$ which is determined by the local velocity gradients $\delta\dot{\mathbf{x}}=(\nabla \mathbf{ u})^\intercal \delta\mathbf{x}$. The $\gamma_T$ are computed according to Eq.~\eqref{gammat} and their statistics are collected for different time intervals $T$. The Lyapunov exponent is obtained as the average $\mean{\gamma_T}=\lambda$. To quantify the effects of the finite resolution on the computation of the spectral correction, we performed an additional set of 15 simulations of Eq.~\eqref{eq1} ($\alpha\in[0.16,2.56]$) at the increased resolution $N=8192$. All simulation parameters are kept constant except from the viscosity, which is reduced to $\nu=5 \times 10^{-6}$.

\begin{table}
\centering
\begin{tabular}{cccccccc||cccccccc}			
    $\alpha$ & $Z$ & $E$ & $\varepsilon_\alpha$ & $\varepsilon_\nu$ & $\eta_\alpha$ & $\eta_\nu$ & $\lambda$ & $\alpha$ & $Z$ & $E$ & $\varepsilon_\alpha$ & $\varepsilon_\nu$ & $\eta_\alpha$ & $\eta_\nu$ & $\lambda$\\
    \hline
    \hline
0.16 & 7.87 & 0.600 & 0.192 & 0.0063 & 2.52 & 0.666 & 0.563\,\, &\, 2.28 & 0.70 & 0.043 & 0.198 & 0.0006 & 3.17 & 0.010 & 0.118 
\\
0.32 & 4.69 & 0.304 & 0.194 & 0.0038 & 3.00 & 0.180 & 0.461\,\, &\, 2.56 & 0.62 & 0.038 & 0.197 & 0.0005 & 3.15 & 0.009 & 0.100 
\\
0.48 & 3.23 & 0.203 & 0.195 & 0.0026 & 3.10 & 0.073 & 0.401\,\, &\, 3.00 & 0.53 & 0.033 & 0.197 & 0.0004 & 3.16 & 0.007 & 0.077 
\\
0.56 & 2.80 & 0.175 & 0.196 & 0.0022 & 3.13 & 0.054 & 0.377\,\, &\, 3.50 & 0.45 & 0.028 & 0.197 & 0.0004 & 3.16 & 0.006 & 0.058 
\\
0.64 & 2.46 & 0.154 & 0.197 & 0.0020 & 3.14 & 0.043 & 0.356\,\, &\, 4.00 & 0.39 & 0.025 & 0.197 & 0.0003 & 3.16 & 0.006 & 0.046 
\\
0.80 & 1.97 & 0.123 & 0.196 & 0.0016 & 3.14 & 0.031 & 0.319\,\, &\, 4.50 & 0.35 & 0.022 & 0.197 & 0.0003 & 3.16 & 0.005 & 0.037 
\\
0.96 & 1.65 & 0.103 & 0.197 & 0.0013 & 3.17 & 0.025 & 0.286\,\, &\, 5.00 & 0.32 & 0.020 & 0.197 & 0.0003 & 3.16 & 0.004 & 0.031 
\\
1.12 & 1.41 & 0.088 & 0.197 & 0.0011 & 3.16 & 0.021 & 0.256\,\, &\, 5.50 & 0.29 & 0.018 & 0.197 & 0.0002 & 3.16 & 0.004 & 0.026 
\\
1.28 & 1.23 & 0.077 & 0.197 & 0.0010 & 3.15 & 0.018 & 0.229\,\, &\, 6.00 & 0.26 & 0.016 & 0.197 & 0.0002 & 3.16 & 0.004 & 0.022 
\\
1.44 & 1.10 & 0.068 & 0.197 & 0.0009 & 3.16 & 0.016 & 0.206\,\, &\, 7.00 & 0.22 & 0.014 & 0.196 & 0.0002 & 3.14 & 0.003 & 0.016 
\\
1.60 & 0.99 & 0.061 & 0.196 & 0.0008 & 3.15 & 0.014 & 0.185\,\, &\, 8.00 & 0.20 & 0.012 & 0.195 & 0.0002 & 3.14 & 0.003 & 0.012 
\\
1.80 & 0.88 & 0.055 & 0.198 & 0.0007 & 3.17 & 0.013 & 0.161\,\, &\, 9.00 & 0.18 & 0.011 & 0.196 & 0.0001 & 3.15 & 0.002 & 0.009 
\\
2.00 & 0.79 & 0.049 & 0.197 & 0.0006 & 3.17 & 0.011 & 0.142\,\, &

\\

    \hline
\end{tabular}
\caption{Simulation parameters. From left to right: friction $\alpha$, time averaged enstrophy $Z$, energy $E$, friction energy dissipation rate $\varepsilon_\alpha$, viscous energy dissipation rate $\varepsilon_\nu$, friction enstrophy dissipation rate $\eta_\alpha$, viscous enstrophy dissipation rate $\eta_\nu$ and Lyapunov exponent $\lambda$. The viscosity $\nu = 4 \times 10^{-4}$ ($\nu=5 \times 10^{-6}$ for the runs at resolution $N=8192$), and the forcing scale defined by $k_I = 4$ and $\delta k = \pm 1$ are the same through all the simulations, while the forcing amplitude is chosen to provide constant energy and enstrophy injection rates $\varepsilon_I = 0.198$ and $\eta_I = 3.185$, respectively.
}
\label{tablegen}
\end{table}

\begin{figure}[h]
\centering
\includegraphics[width=0.8\textwidth]{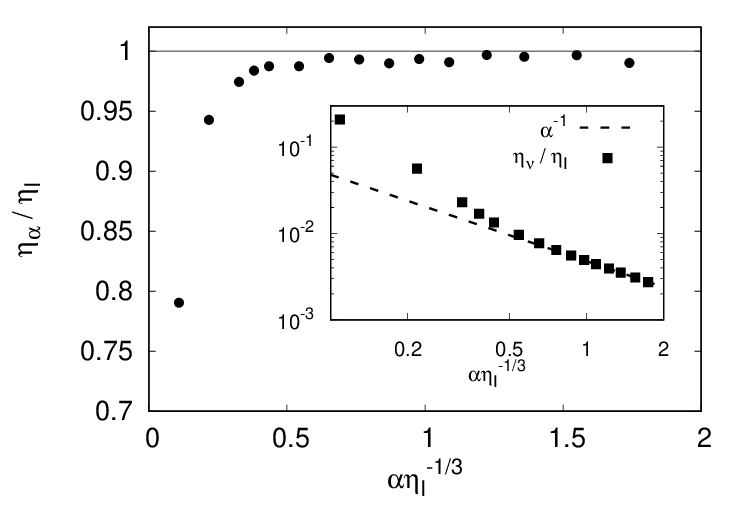}
\caption{Enstrophy dissipation rate due to friction $\eta_{\alpha}$ (main) and viscosity $\eta_{\nu}$ (inset), both normalized by enstrophy input $\eta_{I}$ and shown as functions of nondimensionalised friction.
}
\label{fig1}
\end{figure}

After an initial transient, the flow attains a statistically stationary turbulent regime, in which the energy and enstrophy balances (Eqs.~\eqref{eq2} and \eqref{eq3}) are achieved. The time-averaged values of the energy and enstrophy, as well as the dissipation rates, are reported in Table~\ref{tablegen}. In all cases, the friction force is sufficiently intense to remove completely the energy injected at the forcing scale, as shown by the negligible value of the viscous dissipation of energy $\varepsilon_\nu$ compared with $\varepsilon_\alpha \simeq \varepsilon_I$. 

Focusing on the enstrophy balance, in Figure~\ref{fig1} we show the ratio between the enstrophy dissipation due to friction $\eta_{\alpha}$ and the enstrophy injection rate $\eta_I$ as a function of the friction coefficient $\alpha$. The ratio is close to unity already for moderate friction intensity $\alpha \gtrsim 0.5 \eta_I^{1/3}$, indicating that almost all the injected enstrophy is dissipated by large-scale friction and the viscous dissipation $\eta_\nu$ is negligible. In the inset of Figure~\ref{fig1}, we show that $\eta_\nu$ decreases as $\alpha^{-1}$ at large values of friction, as prescribed by the dimensional estimate of Eq.~\eqref{eq_lim2}. We notice that the scaling $\eta_\nu \sim \alpha^{-1}$ is observed for values of friction $\alpha \gtrsim 0.5 \eta_I^{1/3}$ which produce a spectral correction $\xi(\alpha)>2$, in agreement with the discussion in Sec.~\ref{sec:eqs_2d}. 

The observation that for $\alpha \gtrsim 0.5 \eta_I^{1/3}$ the enstrophy balance is dominated by the friction dissipation $\eta_I \simeq \eta_\alpha = 2\alpha Z$ shows that Eq.~\eqref{eq_lim5} provides an accurate estimate for the enstrophy even at moderate friction intensity.  

\subsection{Distribution of the Finite-Time Lyapunov Exponents}
\label{sec:corr}

\begin{figure}[h]
\centering
\includegraphics[width=0.49\textwidth]{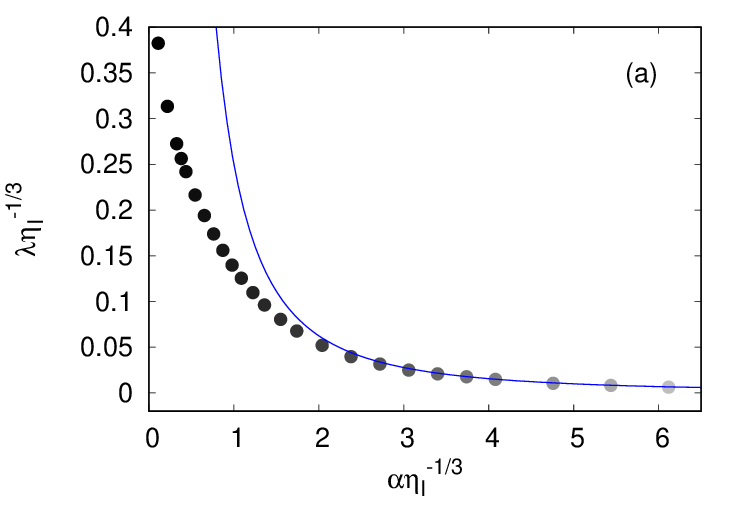}
\includegraphics[width=0.49\textwidth]{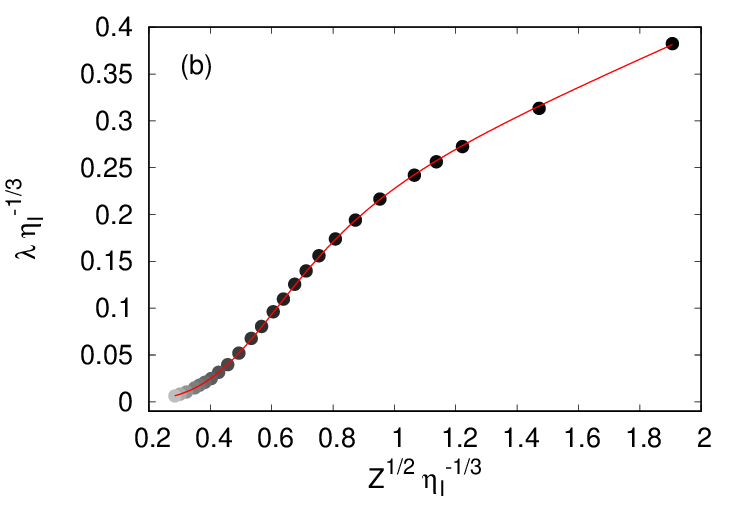}
\caption{(a) Lyapunov exponent as a function of nondimensional friction. The blue line represents the high-friction prediction $\lambda = \eta_I/4\alpha^2$. (b) Lyapunov exponent as a function of the square root of enstrophy (nondimensionalised). The red line shows Eq.~\eqref{model_lyap} with coefficients $A=6.780$, $B=-3.985$ and $C=0.597$.}
\label{fig2}
\end{figure}

In Figure~\ref{fig2}a we show the average Lyapunov exponent $\lambda$ as a function of the friction coefficient $\alpha$. We observe that $\lambda$ is a decreasing function of $\alpha$, showing that increased friction causes a reduction in the chaoticity of the flow. This behavior can indeed be understood from a simple physical argument. As discussed in Sect.~\ref{sec:model_lyap}, the value of the Lyapunov exponent is determined by the intensity of the velocity gradients, which can be quantified by the enstrophy $Z \propto \mean{|\nabla \mathbf{u} |^{2}}_L$, and by their correlation time $\tau$, which is the shortest between the friction time $1/\alpha$ and the advection time $1/\sqrt{Z}$. Recalling that $Z \simeq  \eta_I/2\alpha$, one gets that $\lambda$ always reduces with increasing friction. In the high-friction regime $\alpha \eta_I^{-1/3} \gg 1$ the values of $\lambda$ are in excellent agreement with equation \eqref{eq:lyap_kraichnan} $\lambda = \eta_I/4\alpha^2$. 

The transition from the low-friction regime to the high-friction regime is evident in Figure~\ref{fig2}b, where we show the Lyapunov exponent as a function of $\sqrt{Z}$. 
In the high-friction (low enstrophy) regime, the values of $\lambda$ coincide with the prediction $\lambda = Z^2/\eta_I$. The transition to the linear scaling $\lambda \simeq A\sqrt{Z} - \eta_I^{1/3} B/A^2$ at low-friction (high enstrophy) occurs around $\sqrt{Z}\eta_I^{-1/3} \simeq 1$. 
The generic dependence of $\lambda$ as a function of $\sqrt{Z}$ is perfectly captured by our model~\eqref{model_lyap} for all the values of $Z$. The nondimensional coefficients $A=6.780\pm0.003$, $B=-3.985\pm0.013$, and $C=0.597\pm0.002$ have been obtained by fitting the numerical data. 

To investigate the statistics of the fluctuations of the FTLE, we compute numerically the Cram\'er function from the asymptotic behavior~\eqref{LD} of the probability distribution $P(\gamma_T)$ as follows: 
\begin{equation}
C(\gamma_T) = -\frac{1}{T} \ln\left[\frac{P(\gamma_T)}{P(\lambda)}\right]\ .\
\label{C_num}
\end{equation}
The convergence of (\ref{C_num}) towards the Cram\'er function $C(\gamma)$ is observed for long times $T \gg 1/\lambda$. In particular we find that  $C(\gamma_T)$ becomes independent on $T$ for $T\lambda \gtrsim 5$. This suggests that the suitable time-scales for comparing the Cram\'er functions measured in simulations with different values of the  friction coefficient $\alpha$ 
are proportional to $1/\lambda$. 

\begin{figure}
    \centering
    \includegraphics[width=0.8\textwidth]{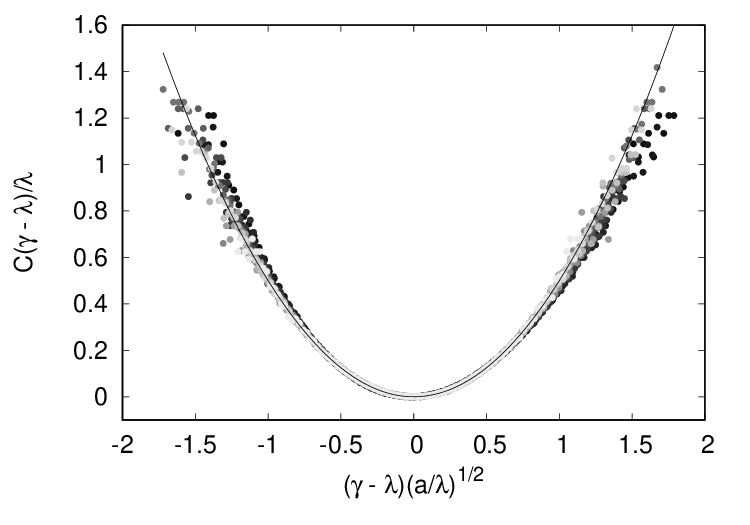}
    \caption{Cram\'er functions for all $\alpha\in[0.16, 2.56]$
    computed at time $T\lambda \simeq 10$. Both axes are normalized to collapse the different curves according to the Gaussian approximation. The darkest points correspond to the lowest friction parameter $\alpha$, and the black line represents $x^2/2$. Note that the Gaussian distribution does not capture the tails for the smaller values of friction (darkest points), indicating a non-negligible skewness.}
        \label{fig3}
\end{figure}

Following this reasoning, in Figure~\ref{fig3}, we show the rescaled Cram\'er functions $C/\lambda$ at times $T\lambda \simeq 10$ obtained in simulations with different friction coefficient $\alpha$. It is natural to rescale the argument with the standard deviation of $P(\gamma_T)$ expressed in terms of the parameter $a$ (see \eqref{limit2}) as $1/\sqrt{aT}$. At fixed $T\lambda$ the rescaled argument becomes $(\gamma_T-\lambda)\sqrt{a/\lambda}$. 
The collapse of the curves in Figure~\ref{fig3} shows that all Cram\'er functions are well approximated by the quadratic form \eqref{C2} for all the values of $\alpha$, in particular for large friction as predicted by Eq.~\eqref{eq:Cramer_kraichnan}.

\begin{figure}[t]
\centering
\includegraphics[width=0.8\textwidth]{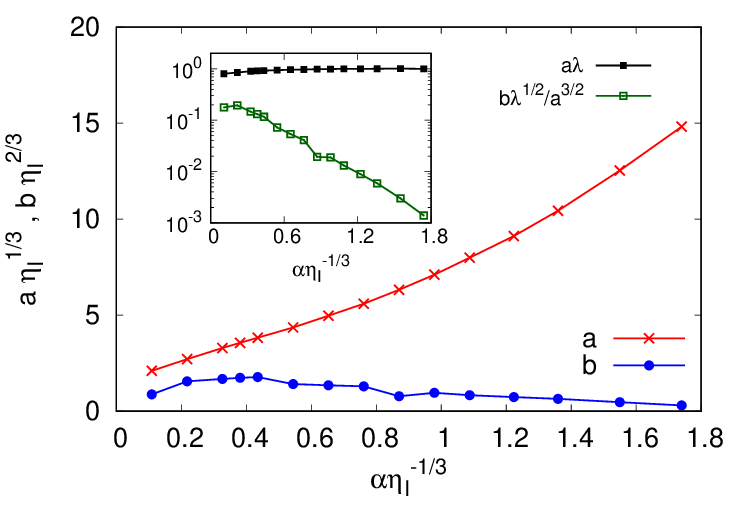}
\caption{Friction dependence of the Cram\'er functions' parameter $a$ and $b$.}
\label{fig4}
\end{figure}

In Figure~\ref{fig4}, we show the dependence of the parameter $a$ on the friction coefficient. The monotonic growth of $a$ as a function of $\alpha$ reveals that the variance of $\gamma_T$ decreases with increasing friction. Notably, we find that $a\lambda \to 1$ for $\alpha \gtrsim 0.5 \eta_I^{1/3}$ (see inset of Fig.~\ref{fig4}), in agreement with the high-fricton predictions (see Sect.~\ref{sec:model_lyap}).


A more in-depth analysis of the Cram\'er functions reveals the presence of a weak asymmetry, which is more pronounced in the cases with low friction. A quantitative measure of the asymmetry of the $C(\gamma)$ is provided by the skewness of the distribution of $\gamma_T$. Within the third-order polynomial approximation~\eqref{C3} of $C(\gamma)$, the skewness is given by
\begin{equation}\label{skew}
    \frac{\mean{\left(\gamma_T-\lambda\right)^3}}{\mean{\left(\gamma_T-\lambda\right)^2}^{3/2}}=
    \frac{2b\lambda^{1/2}}{a^{3/2}(T\lambda)^{1/2}}+\mathcal{O}\left((T\lambda)^{-1}\right).
\end{equation}
Equation~\eqref{skew} states that the skewness of $P(\gamma_T)$ decreases with $T$ as $T^{-1/2}$. 

The values of the parameter $b$ have been determined according to Eq.~\eqref{limit3} and they are reported in Figure~\ref{fig4} as a function of $\alpha$. They show a weak, non-monotonic dependence on $\alpha$, with a maximum around $\alpha\eta_I^{-1/3} \approx 0.4$. For values $\alpha \gtrsim 0.4 \eta_I^{1/3}$, 
we observe a decrease of $b$ as a function of $\alpha$, in agreement with the discussion in Sect.~\ref{sec:model_lyap}. 

From the dependence of $\lambda$, $a$, and $b$ on $\alpha$, we obtain that the factor  $b\lambda^{1/2}/a^{3/2}$, which is proportional to the skewness at fixed $T\lambda$, decreases rapidly (almost exponentially) as a function of the friction intensity (see inset of Fig.~\ref{fig4}). In summary, for fixed $T\lambda$, larger friction produces an FTLE distribution closer to Gaussian, while, in the same limit, the size of the typical fluctuations around $\lambda$ is given by $\sqrt{1/Ta} \simeq \lambda $.

\subsection{Spectral correction}\label{PARCORR}

\begin{figure}[h!]
\centering
\includegraphics[width=0.49\textwidth]{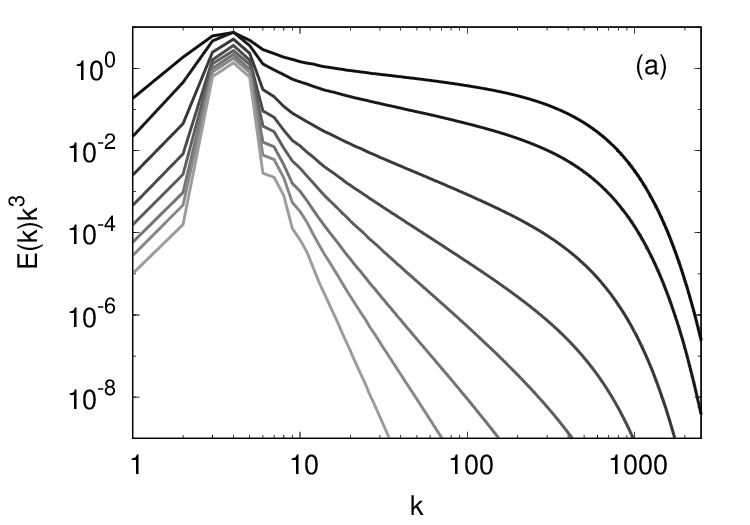}
\includegraphics[width=0.49\textwidth]{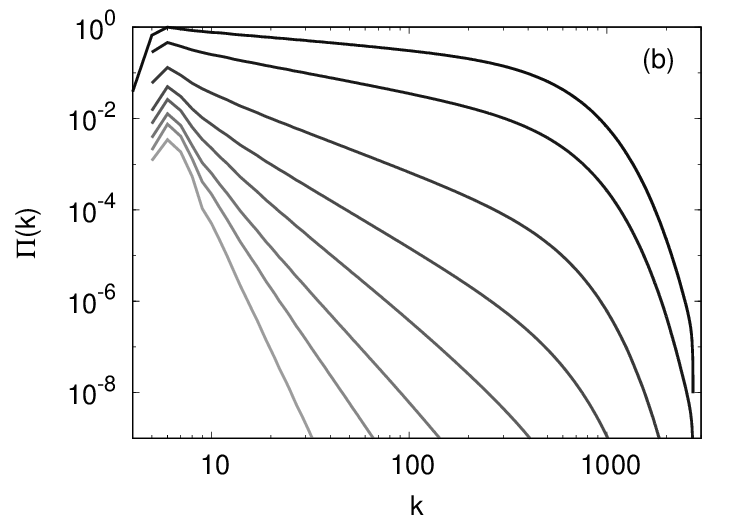}
\caption{(a) Energy spectra $E(k)$ compensated with the dimensional scaling $k^3$ and (b) corresponding enstrophy flux $\Pi(k)$ for $\alpha=[0.16, 0.32, 0.64, 0.96, 1.28, 1.60, 2.00, 2.56]$ (from top to bottom). Simulations at resolution $N=8192$.}
\label{fig5}
\end{figure}

In Figure~\ref{fig5}, we show the energy spectra for the different values of the friction coefficient. As expected, we observe a power-law scaling at intermediate wavenumbers where $E(k)\sim k^{-(3+\xi)}$. To determine precisely the spectral correction, we measured the exponent $\xi$ from the power-law scaling of the enstrophy flux $\Pi_Z(k)\sim k^{-\xi}$ using the standard logarithmic derivative method. As discussed in detail in \cite{Valadao2025}, this procedure gives a more robust evaluation of $\xi$, in particular in the case of small friction. Nonetheless, we found that the simulations at resolution $N=1024$ do not allow a precise determination of the exponent $\xi$ since its value is affected by a relatively large uncertainty which depends on the range of wavenumbers on which it is measured. To have a more accurate measure of $\xi$, we therefore used the simulations at resolution $N=8192$.

\begin{figure}[h!]  
\centering
\includegraphics[width=0.8\textwidth]{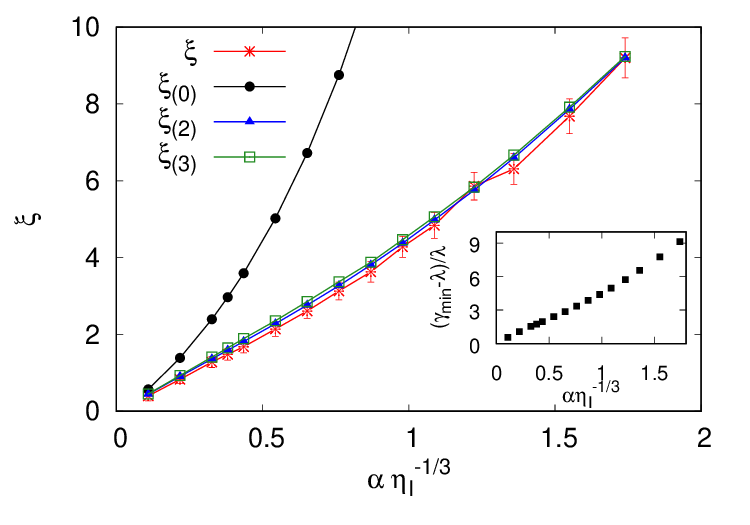}
\caption{Spectral correction $\xi$ measured from the enstrophy flux (red crosses with errorbars) compared with the mean-field prediction $\xi_{(0)}=2 \alpha/\lambda$ (black circles) and the ones based on the quadratic (\ref{chi2}) ($\xi_{(2)}$, blue triangles) and cubic  ($\xi_{(3)}$, green squares) approximations of the Cram\'er function. 
Inset: relative distance between $\gamma_{\min}$ in the quadratic approximation (\ref{gamma-min}) and the Lyapunov exponent $\lambda$.} 
\label{fig6}
\end{figure}

Figure~\ref{fig6} shows the measured spectral correction $\xi$ together with the different predictions obtained from the Cram\'er function. It is evident that the mean-field approximation $\xi_{(0)}$ is very far from the numerical data and that the discrepancy increases dramatically for large friction. On the contrary, the exponent $\xi_{(2)}$ based on the quadratic approximation of the Cram\'er function (\ref{chi2}) is very close to the numerical results in the whole range of $\alpha$. 

It is interesting to observe that the value of $\gamma$ which minimizes (\ref{chi}) in the quadratic approximation is given by
\begin{equation}
\gamma_{\min}=\lambda\sqrt{1+\frac{4\alpha}{a\lambda^2}}\ ,\
\label{gamma-min}
\end{equation}
which shows that $\gamma_{\min}\geq\lambda$. Moreover, since (\ref{chi2}) can be written as $\xi_{(2)}=a(\gamma_{\min}-\lambda)$, we see that the exponent correction is proportional to the distance between $\lambda$ and the value $\gamma_{\min}$ of the fluctuation selected by the steepest descent argument.
As the friction increases, $\gamma_{\min}$ moves away from $\lambda$ (see the inset of Fig.~\ref{fig6}) and the width of the Cram\'er function shrinks as $1/a\rightarrow0$ such that the mean field approximation $(\gamma_{\min}=\lambda)$ becomes worse.
Figure~\ref{fig6} also shows the scaling exponent $\xi_{(3)}$ obtained from the cubic approximation of the Cram\'er function. Since the skewness of the Cram\'er function is very small, the additional correction to $\xi$ induced by the cubic term is  negligible.

\section{Conclusions}\label{sec:concl}

In this work, we have reported a theoretical and numerical investigation of the influence of linear friction on Lagrangian chaos in two-dimensional turbulent flows. The chaotic nature of the Lagrangian trajectories has been quantified in terms of the statistics of finite-time Lyapunov exponents. 

Our results demonstrate that an increasing friction causes a reduction of Lagrangian chaos, as shown by the decrease of the leading Lyapunov exponent $\lambda$ as a function of $\alpha$. The analysis of the relation between $\lambda$ and the enstrophy $Z$ reveals the presence of two distinct regimes, at low and high friction, respectively. In the regime of high friction, we derived theoretical predictions based on the Kraichnan ensemble for the dependence of the Lyapunov exponent and the Cramér functions on the enstrophy and friction intensity. Furthermore, we propose a general model which describes the transition from the high-friction regime to the low-friction one, providing an explicit expression for $\lambda$, a small-scale Lagrangian observable, as a function of the enstrophy, a large-scale global observable. 
  
The FTLE fluctuations around their mean value $\lambda$ have been investigated in terms of the Cram\'er function within the framework of the large-deviations theory. At high friction, we found that the Cramér function is well approximated by a quadratic function, i.e., the fluctuations of the FTLE correspond to Gaussian statistics. The coefficient $a$ of the quadratic approximation is found to be inversely proportional to $\lambda$. This implies that the statistics of the FTLE are completely determined by its mean value $\lambda$. At low friction, we observe a weak asymmetry in the Cramér functions, which corresponds to a positive skewness of the distibution, indicating a slightly higher probability of observing a positive fluctuation of the FTLE. 

Finally, we addressed the steepening of the energy spectrum in the range of wavenumbers of the direct enstrophy cascade, resulting from the combined effect of linear friction with chaotic stretching of vorticity filaments at small scales. The spectral correction $\xi$ measured in the numerical simulations is found to be in good agreement with the theoretical prediction obtained in terms of the statistics of the FTLE. In particular, both the quadratic and cubic approximations of the Cram\'er function provide reliable estimates of $\xi$. On the contrary, the prediction obtained from a mean-field approximation is not accurate. This highlights the importance of taking into account the statistical fluctuations of the FLTE. 

Our findings reveal that, in the presence of strong linear friction, the
statistics of the FTLE in two-dimensional turbulence are mostly determined by
the dynamics at the forcing scale. This suggests the possibility that the
statistics could depend non-trivially on the details of the external forcing.
It would be interesting to systematically investigate different forcing
protocols in future studies. On a different note, it would also be interesting to
apply the same approach to different systems. The hypotheses of the theoretical
framework used here seem quite general, so in principle they should be applicable to other equations.
However, we are not aware of any other physical system displaying both the necessary scaling properties and the correct dynamics.
Magneto-hydrodynamic turbulence for example, has an advective dynamics (which
is necessary for the connection with the Lagrangian FTLE) but it does not exibit a
smooth direct
cascade regime \cite{biskamp_1993}. Wave turbulence models \cite{nazarenko2011wave}, on the other hand, are not in general dominated by an advective non-linear term driving the downscale transfer of the cascading quantity. Although we cannot
exclude that similar approaches can be used for turbulent wave equations or closely related quantum turbulence models (e.g., the Gross-Pitaevskii model \cite{zhu2023direct}), the Lagrangian side of the theoretical argument seems hard to justify.

\section*{Acknowledgement}

We acknowledge HPC CINECA for computing resources within the INFN-CINECA Grants INFN24-FieldTurb and INFN25-FieldTurb. This work was supported by Italian Research Center on High Performance Computing Big Data and Quantum Computing (ICSC), project funded by European Union - NextGenerationEU - and National Recovery and Resilience Plan (NRRP) - Mission 4 Component 2 within the activities of Spoke 3 (Astrophysics and Cosmos Observations).



\end{document}